\documentclass[twocolumn]{aa}
\usepackage{graphicx}
\usepackage{txfonts}
\begin{document}

\newcommand {\flux} {{$\times$ 10$^{-11}$ erg cm$^{-2}$ s$^{-1}$}}
\newcommand {\eg}  {3EG J2027+3429 }
\newcommand {\wga} {WGA J2025.1+3342 }
\def\magcir{\raise -2.truept\hbox{\rlap{\hbox{$\sim$}}\raise5.truept
\hbox{$>$}\ }}

\title{3EG J2027+3429 another Blazar behind the Galactic Plane}

\titlerunning{3EG J2027+3429 another Blazar behind the Galactic Plane}

\author{V. Sguera\inst{1}, A. Malizia\inst{2}, L. Bassani\inst{2}, J. B. Stephen\inst{2},  
           G. Di Cocco\inst{2}} 
        
\authorrunning{Sguera et al.}

\offprints{sguera@bo.iasf.cnr.it}

\institute{Dipartimento di Astronomia di Bologna, via Ranzani 1, I-40127 Bologna, Italy
 	 \and 
           Istituto di Astrofisica Spaziale e Fisica Cosmica – Sezione di Bologna, 
           Area di Ricerca di Bologna, Via P. Gobetti 101, Bologna, ITALY}

    \date{Received / accepted}

\abstract{We report on the association of an X-ray source (WGA J2025.1+3342), serendipitously
found with BeppoSAX in two separate observations, with the
unidentified EGRET source 3EG J2027+3429.  The source is detected from
1 keV up to about 100 keV, has a flat ($\Gamma$=0.6-1.5) spectrum and
is highly variable both in intensity and shape. The data indicate
marginal evidence for an iron line in the source rest frame. The
overall X-ray luminosity is ${\sim}$ 4 10$^{45}$ erg s$^{-1}$ typical
of a quasar.  The X-ray source is coincident in radio with a bright
object characterized by a flat spectrum over the band 0.3-10 GHz while
in optical it is identified with a quasar at redshift 0.22. All
available data indicate a SED compatible with a low frequency peaked
or red blazar type object.  This identification is interesting because
this is the second blazar found behind the galactic plane in the
direction of the Cygnus region.  }


\maketitle


\section{Introduction}

The nature of unidentified gamma-ray sources has remained a mystery
since the first survey of the gamma-ray
sky with the COS-B satellite.  More than sixty percent of the 271 high
energy sources reported in the 3rd EGRET Catalogue (Hartman
et al. 1999) are still unidentified, with no firmly established counterparts
at other wavebands.  Considering their distribution on the sky, we can
infer that one third of these are extragalactic (probably AGNs of the blazar type)
the rest most likely being objects within the Milky Way. Most of
the AGN blazars are strong flat spectrum radio sources (i.e. jet
dominated) and are variable in the gamma-ray band.  Of the galactic
population, steady sources are likely to be radio quiet pulsars, while
transient objects are still poorly understood and may be due to
interactions of individual pulsars or neutron star binaries with the
ambient interstellar medium.  The main difficulty in identifying
EGRET sources is their often large error
box; therefore a positional correlation with a known object is usually
not enough to identify a source. For this reason, a multiwavelength
approach, using X-ray, optical and radio data, is often needed to
understand the nature of these sources (see Caraveo 2002 and references therein).  
Searches for X-ray
counterparts, especially at high energies, are particularly useful
in finding a positionally-correlated, highly unusual object with
special parameters that might be expected to produce gamma rays. 
For this reason we have recently started a program to search the
BeppoSAX archive for observations covering either partially or totally EGRET
error boxes (Sguera et al. 2003).  During this search, we
discovered a hard X-ray source within the 99{\%} confidence
position of 3EG J2027+3429 (also known as 2EG J2026+3610). The
gamma-ray source is at a low galactic latitude in the Cygnus region
and has recently been proposed as being associated with an extragalactic
object on the basis of a radio search and optical follow up
observation (Sowards-Emmerd et al. 2002). Previous works have suggested a
young pulsar as a possible counterpart, given the presence in the
EGRET error box of an OB star association (Zhang and Cheng
2000, Romero, Benaglia and Torres 1999) although the gamma-ray flux
variability and spectral characteristics (Merck et al. 1996)
argue against this possibility.  The extragalactic interpretation is
instead more attractive given the nature of the proposed optical
counterpart (a Flat Spectrum Radio Quasar or FSRQ) and variability of the gamma-ray
emission, although this would be unusual given the proximity of the source to
the Galactic Plane. We show in the present paper that the
extragalactic optical/radio source is also an X-ray emitter from 
0.2 keV to $\sim$100 keV and furthermore that it displays variability
at these frequencies.  Our X-ray data together with an overall
assessment of the source's spectral  characteristics confirm the blazar
nature of this object and further indicate that now 2 such sources
have been found at low galactic latitudes after the identification of  3EG J2016+3657  
as a blazar-like radio source G74.87+1.22 (Mukherjee et al. 2000, Halpern et
al. 2001) .

\begin{figure}
\centering
\includegraphics[width=8.0cm,height=8.0cm]{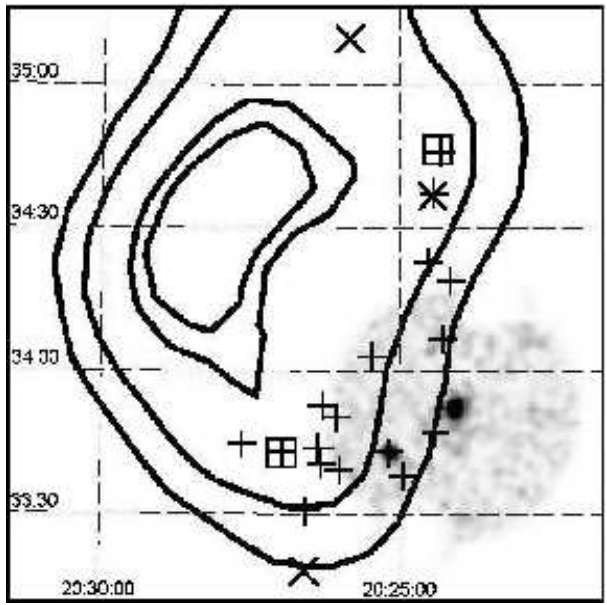}
\caption{The BeppoSAX/MECS (2-10 keV) image superimposed on the EGRET $\gamma$-ray 
probability contours at 50\%, 68\%, 95\%, 99\% confidence level. All the X-ray
counterparts listed in table 1 are also shown: pluses  are WGA sources,
crosses are ROSAT faint sources and squares are the ROSAT bright sources.}
\end{figure}

\section{Search for X-ray counterparts in the EGRET error box}

The X-ray counterpart of 3EG J2027+3429 was found in two BeppoSAX-MECS
observations targeted at the black hole candidate GS2023+338 (also
V404 Cygni).  In figure 1, one of these two BeppoSAX-MECS (2-10 keV)
images is superimposed on the EGRET $\gamma$-ray probability contours
(50\%, 68\%, 95\% and 99\%). The black hole candidate lies just outside
these contours which makes it less likely to be the X-ray
counterpart of the EGRET source. Instead, another source, also well visible in the 2-10
keV image, is located within the 99{\%} contour at (J2000) RA = 20h
25m 0.7s Dec = +33$^{\circ}$ 42' 52'' , 25 arcmin north-east of the target source
(the uncertainty associated with the source position is
1 arcmin, 90\% confidence level). The source is visible in both MECS
observations but is only marginally detected in the only available
LECS image, thus indicating a hard X-ray spectrum.\\ GS2023+338 was
also observed by the ASCA satellite but unfortunately the EGRET
candidate counterpart is at the edge of the GIS telescope field of view and although
it is well detected, the data are too affected by vignetting to
provide useful information on its spectral shape (HEASARC
database).\\
There is also a pointed ROSAT observation of the black hole candidate,
which provides a total of 17 X-ray sources reported in the WGA
catalogue (White, Giommi and Angelini 1994).  Furthermore, the cross correlation of the 99 {\%}
EGRET error box with the ROSAT All-Sky-Survey catalogues (Bright and
Faint) resulted in 5 objects, 3 of which are also WGA
sources. Therefore all together there are 19 X-ray ROSAT sources within the
EGRET error box. All these soft X-ray candidates are listed in Table 1
together with their coordinates, ROSAT count rates and 
offset from the 3EG source position.  Most of these objects are
unidentified in the Simbad/NED databases except for 6 objects which are
associated with normal stars (sources n.4, n.9, n.15, n.19, n.20).  
Only one source, n. 13 (WGA J2025.1+3342) in table 1, is detected at energies greater than a few keV and
it is the source coincident with the object serendipitously detected by BeppoSAX
and ASCA.  The ROSAT count rate of WGA n.13 provides a crude flux estimate of
$\sim$10$^{-13}$ erg cm$^{-2}$ s$^{-1}$ in the 0.05-2 keV band, compatible
with the marginal detection by BeppoSAX at low energies given the low
exposure of the LECS instrument.\\ 
A cross correlation of all X-ray objects with radio catalogues available from the HEASARC
database indicates that only 2 objects are radio emitting: WGA n.11
(also RXS n.22) and WGA n.13. These radio sources have a 20 cm flux of 5
and 1268 mJy respectively and therefore the first object is not
sufficiently bright to be a likely counterpart of a  gamma-ray source
(see section 4).  All other radio sources detected within the EGRET
99{\%} error contour are also all too faint (all their fluxes are below 200
mJy at 1.5 and 5 GHz) to be likely associations of the gamma-ray
source.\\ To conclude, WGA n.13 is not only the strongest hard X-ray
source in the EGRET error box but it is also the brightest radio
object. In optical, this source was recently observed by
Sowards-Emmerd et al. (2002) who obtained the first spectrum and
measured a redshift of z=0.22.  The spectrum shows emission lines of
the Balmer series and so the source can be optically classified as a Quasar.\\

\begin{table*}
\begin{center}
\caption{ROSAT sources in the EGRET error box.}
\begin{tabular}{llcccccc}
\hline
 & Source                   & RA            & Dec          & Count Rate   & Search Offset   & Type   & Radio Count. \\
 &                          & (J2000)       & J(2000)      & Cts/s        & $arcmin$          &        &              \\
\hline
1  & 1WGA J2024.4+3437      & 20 24 25.90    & +34 37 29.0    & 0.0135    &  31.542            &  -     &  No \\  
2  & 1WGA J2024.5+3422      & 20 24 31.32    & +34 22 44.0    & 0.0060    &  31.854            &  -     &  No \\
3  & 1WGA J2024.3+3445      & 20 24 18.90    & +34 45 48.0    & 0.0275    &  35.057            &  -     &  No \\
4  & 1WGA J2025.4+3403      & 20 25 28.40    & +34 03 06.0    & 0.0017    &  35.119            &  -     &  No \\
5  & 1WGA J2024.1+3418      & 20 24 08.20    & +34 18 50.0    & 0.0024    &  37.692            &  -     &  No \\
6  & 1WGA J2026.2+3352      & 20 26 17.30    & +33 52 52.0    & 0.0022    &  40.982            &  -     &  No \\
7  & 1WGA J2024.2+3406      & 20 24 15.90    & +34 06 40.0    & 0.0046    &  42.523            &  -     &  No \\
8  & 1WGA J2026.0+3350      & 20 26 03.60    & +33 50 24.0    & 0.0032    &  44.041            &  -     &  No \\
9  & 1WGA J2027.6+3344      & 20 27 38.00    & +33 44 41.0    & 0.0080    &  49.036            &  -     &  No \\
10 & 1WGA J2026.3+3343      & 20 26 21.20    & +33 43 54.0    & 0.0056    &  49.676            &  -     &  No \\
11 & 1WGA J2026.9+3343      & 20 26 57.50    & +33 43 08.0    & 0.0918    &  49.873            &  -     &  Yes \\
12 & 1WGA J2026.3+3340      & 20 26 19.00    & +33 40 50.0    & 0.0066    &  52.775            &  -     &  No\\
13 & 1WGA J2025.1+3342      & 20 25 10.50    & +33 43 00.0    & 0.0070    &  54.691            &  -     &  Yes \\
14 & 1WGA J2025.9+3339      & 20 25 59.60    & +33 39 14.0    & 0.0089    &  55.088            &  -     &  No\\
15 & 1WGA J2024.3+3347      & 20 24 23.50    & +33 47 16.0    & 0.0009    &  55.741            &  -     &  No \\
16 & 1WGA J2024.9+3338      & 20 24 55.00    & +33 38 05.0    & 0.0016    &  60.495            &  -     &  No \\
17 & 1WGA J2026.5+3330      & 20 26 35.00    & +33 30 39.0    & 0.0098    &  62.526            &  -     & No \\    
18 & 1RXS J202427.2+343641$^{(a)}$  & 20 24 27.19    & +34 36 41.0    & 0.0215    &  31.177            & F      & No \\ 
19 & 1RXS J202550.8+350938  & 20 25 50.80    & +35 09 38.5    & 0.0144    &  39.120            & F      & No \\ 
20 & 1RXS J202635.2+331758  & 20 26 35.20    & +33 17 58.5    & 0.0168    &  75.169            & F      & No \\
21 & 1RXS J202422.2+344631$^{(b)}$  & 20 24 22.20    & +34 46 31.5    & 0.0500      &  34.699            & B      & No\\
22 & 1RXS J202658.5+334253$^{(c)}$  & 20 26 58.50   & +33 42 53.0  & 0.0600      &  50.117    & B      & Yes\\
\hline
\hline
\end{tabular}
\end{center}
Note: ($^{a}$) = also a WGA source: 1WGA J2024.4+3437 (source n. 1 in the table), 
($^{b}$) = also a WGA source: 1WGA J2025.4+3445 (n. 3 in the table),  
($^{c}$) = also a WGA source: 1WGA J2026.9+3343 (n. 11 in the table). 
\end{table*}

\section{BeppoSAX observations of WGA J2025.1+3342}

BeppoSAX-NFI pointed at GS 2023+338 in
September 1996 and December 2000, for an effective MECS exposure time of
21  and 48 ks respectively;  no LECS data are available for
the first observation while the second has an exposure of only 6 ks. 
The PDS exposures are instead 10  and 21 ks for the first and second pointings.
Standard data reduction was performed on the BeppoSAX data using the
software package "SAXDAS" (Fiore, Guainazzi \& Grandi 1998).  
The reduction procedures and screening criteria used to
produce the linearized and equalized (between the two MECS) event
files were standard and took into account the
offset position of the source.  For the PDS data we adopted a fine
energy and temperature dependent Rise Time selection, which decreases
the PDS background by $\sim 40 \%$, improving the signal to noise
ratio of faint sources by about 1.5 (Frontera et al. 1997).  Spectral
fits were performed using the XSPEC 11.0.1 software package and public
response matrices for the off-axis sources as from the 1998 November
issue.  PI (Pulse Invariant) channels were rebinned in order  to
sample the instrument resolution with the same number of channels at
all energies when possible and to have at least 20 counts per bin.
This allows the use of the $\chi^2$ method in determining the best fit
parameters, since the distribution in each channel can be considered
Gaussian. In the following analysis, we use an
absorbed component to take into account the high galactic
column density that in this direction is 7.7 $\times$ 10$^{21}$ cm$^{-2}$ 
(obtained from 21 cm radio data provided by XSPEC). 
The quoted errors correspond to
90\% confidence level for one interesting parameter ($\Delta\chi^2$ =
2.71). The X-ray
candidate for the \eg source is well detected in the 2-10 keV energy
range at 13$\sigma$ and 9$\sigma$ level  but  it is barely visible in the only 
available LECS image. At higher energy, in the
PDS regime, a 4$\sigma$ detection was measured only in the second
observation.\\ 
The PDS instrument has no imaging capability and a field
of view of 1.3$^{\circ}$ (FWHM), hexagonal in shape. 
Therefore any object within this
field of view can be responsible for the high energy emission. As a
first  step it is  natural to attribute the PDS emission to the black
hole candidate; we therefore checked if this is indeed true by
performing a spectral analysis of the combined MECS data of GS2023+338
with high energy PDS data. The source was quiescent during the 2000
BeppoSAX observation and its  X-ray  spectrum was well represented
by a power law ($\Gamma$ $\sim$1.9) absorbed by the  galactic column density
(Campana et al. 2001).
A fit with this model to the combined MECS/PDS data sets
provides a value of 15$^{+11}_{-8}$ for the cross calibration constant 
between the two instruments; this value is well outside
the nominal range of 0.75-0.95 reported by Fiore, Guainazzi \& Grandi
(1998). Furthermore, extrapolating into
the PDS (20-100 keV) energy band,  the low energy power law gives a flux of 6.1 $\times$ 10$^{-13}$ erg cm$^{-2}$ s$^{-1}$ 
which is below the measured PDS flux. Both these facts strongly indicate 
that the PDS emission either is 
not due to the black hole candidate or is at least  highly
contaminated by the presence of another X-ray source in the PDS field
of view. We have searched the HEASARC data base for other possible 
high energy emitting sources located within the PDS field of view but
have found none   that can be considered a likely hard X-ray emitter.\\ 
We therefore assume that the high
energy emission, or at least the major part of it,  comes from the
X-ray counterpart of the \eg object; this is reasonable given the
hard X-ray spectrum of the source compared to the black hole candidate (see below).  
Consequently we fitted
contemporaneously the MECS data from  this new  source  and the  PDS points as performed for
GS2023+338.\\
The 2-10 keV spectrum of the first observation is  simply
fitted with a power law of photon index
$\Gamma$=1.41$^{+0.45}_{-0.41}$ (once galactic absorption is accounted for)  
and provides a 2-10 keV unabsorbed flux of $\sim$3 $\times$ 10$^{-12}$ erg cm$^{-2}$ s$^{-1}$ and 
a corresponding luminosity (at the reported redshift and assuming H$_{\circ}$= 50, q$_{\circ}$= 0)
of $\sim$7 $\times$ 10$^{44}$ erg s$^{-1}$.  
We then concentrated on the second observation and first checked the MECS/PDS 
cross calibration constant which turned out to be
1.5$\pm$1.1, compatible within the errors with the nominal range expected.
Therefore, in the following analysis we impose that this constant varies between 0.75-0.95, i.e. 
within the expected range of values.  The broad band (1-100 keV) spectrum
of the source provides for the second observation  a hard power
law with photon index of $\Gamma$=0.63$^{+0.21}_{-0.25}$; furthermore
marginal evidence for a line feature at around 5 keV is also visible
in the residual of the data to model ratio.  When a Gaussian line in
the rest frame of the source ($z$=0.22) is added to the power law
model, the fit slightly improves. The line is required by the data
only at 72\% confidence level using the F-test, is centered at
E$_{line}\sim$ 6 keV and has  an equivalent width  EW$\sim$
780$^{+621}_{-580}$ eV.  The best fit of the broad band spectrum
during this second observation  is shown in figure 2. The fluxes corrected 
for galactic absorption are 7.4  $\times$ 10$^{-13}$  and 1.6 $\times$ 10$^{-11}$ erg cm$^{-2}$ s$^{-1}$ for
the 2-10 and 20-100 keV band respectively; the corresponding luminosities at the reported redshift 
are  1.4 $\times$ 10$^{44}$ and 3.1 $\times$ 10$^{45}$  erg s$^{-1}$.  
We therefore conclude that the source is highly
luminous in X-rays and variable both in shape and
intensity at these energies.

\begin{figure}
\centering
\includegraphics[width=8cm,height=8cm,angle=-90]{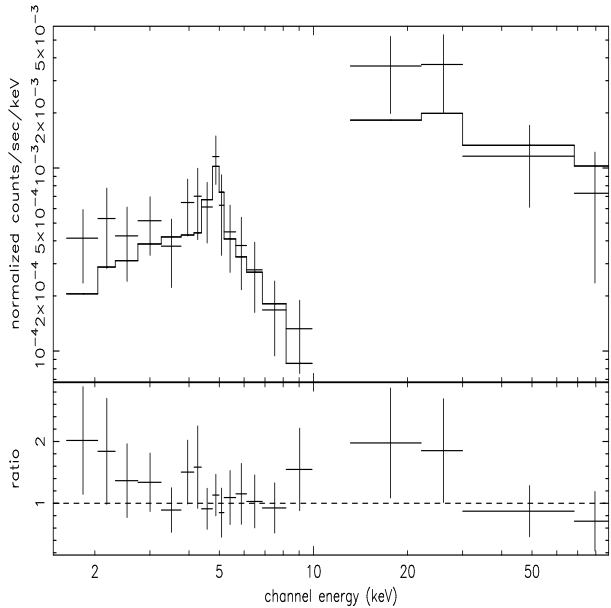}
\caption{BeppoSAX broad band (1-100) keV of \wga during 
the second observation.}
\end{figure}

\section{Source characteristics}

In this section we analyze the broad band spectral characteristics of
the source  WGA J2025.1+3342 which we propose as the X-ray counterpart of 3EG
J2027+3429 and show that it is indeed a new blazar gamma-ray source
behind the Galactic Plane.  The EGRET gamma-ray spectrum is characterized by a power law
photon index  $\Gamma$=2.28 $^{+0.15}_{-0.15}$ close to the average photon index of 2.2 found
in gamma-ray blazars (Mukherjee 2001). Furthermore, the EGRET light curve from the
beginning to the end of the mission, obtained from the 3EG catologue (Hartman et al. 1999),
clearly indicates variability of the gamma-ray flux (see figure 3). This was also noticed by
McLaughin et al. (1996) who listed  3EG J2027+3429 as one of the galactic plane variable
sources. Applying a $\chi^2$  test to the light curve yields a
variability index of 1.19 which corresponds to a probability of
$\sim$ 0.065 that these data are produced by an intrinsically non variable
source (McLaughin et al. 1996). The other possible gamma-ray candidate at these low galactic
latitudes would be a pulsar, but this association is excluded by the measured variability.
Our proposed counterpart is a strong radio source detected from
0.33 to 10 GHz (see table 2 for details). From these broad band radio
data we can estimate a power law energy index ${\alpha}$ =-0.3 which
confirms the flat spectral nature of the source (see also NED where the
radio index is reported to be -0.2); since FSRQ must have
${\alpha}$ ${\le}$0.5 ($F(\nu)\propto\nu^{-\alpha}$) our
source is obviously of this type.  

\begin{table*}
\begin{center}
\caption{Radio sources associated with 1WGA J2025.1+3342}
\begin{tabular}{llcclcc}
\hline
    & Source                     & RA            & Dec           & Flux              & Search Offset    & Radio Catalog                 \\
    &                            & (J2000)       & J(2000)       & mJy               & arcmin                                           \\
\hline
 1  & NVSS J202510+334300        & 20 25 10.80   & +33 43 00.00  & 1268 (20 cm)      & 0.07             & NRAO VLA sky survey catalog    \\
 2  & 2023+3333                  & 20 25 09.85   & +33 43 22.60  & 2104 (6~ cm)       & 0.40             & 6cm Radio Catalog              \\
 3  & WN 2023.2+3333             & 20 25 10.85   & +33 43 01.40  & 982~  (92 cm)      & 0.07             & Westerbork northern sky survey \\
 4  & TXS 2023+335               & 20 25 10.84   & +33 43 00.93  & 1121 (82 cm)      & 0.07             & Texas Survey at 365 MHz \\
 5  & J2025+3343                 & 20 25 10.94   & +33 43 00.21  & 2728 (3.5cm)      & 0.09             & CLASS Survey at 8.4 GHz \\
\hline
\hline
\end{tabular}
\end{center}
\end{table*}

\begin{figure}
\centering
\includegraphics[width=9cm,height=7cm]{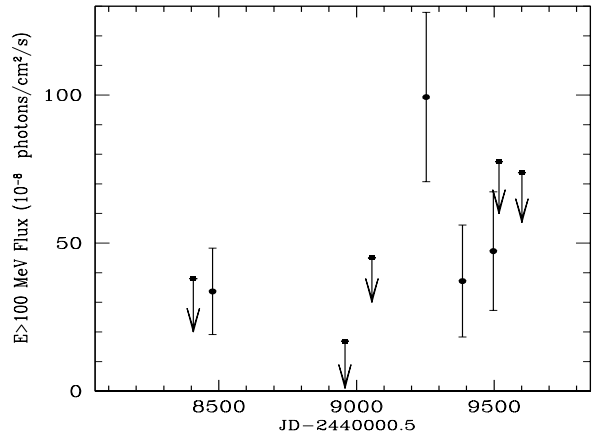}
\caption{Light curve of \eg from May 1991 to Jul 1994.}
\end{figure}

\begin{figure}
\centering
\includegraphics[width=9cm,height=7cm]{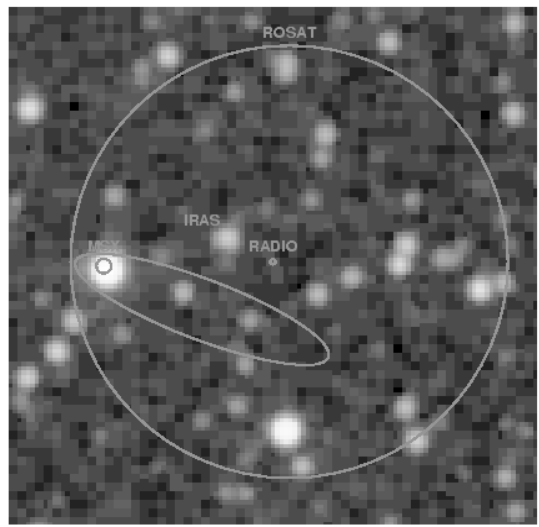}
\caption{2MASS (J band) image.}
\end{figure}

\begin{figure}
\centering
\includegraphics[width=9cm,height=7cm]{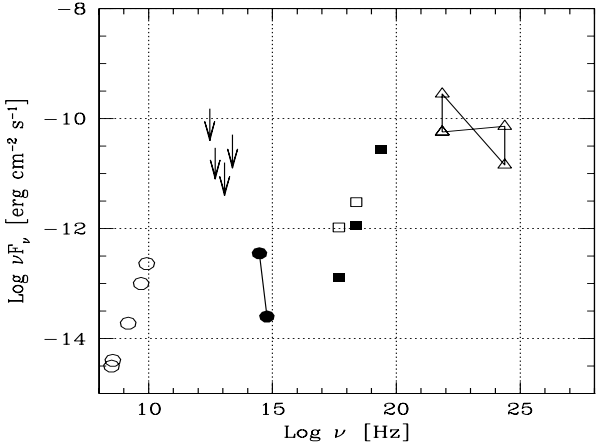}
\caption{Spectral Energy Distribution (SED) of 3EG J2027+3429. Open circles are radio measurements, filled 
circles are optical measurements, squares are our BeppoSAX observation (open squared OBS1 and filled  
square OBS2) and triangles are Egret measurements. The arrows indicate IRAS upper limits at 12$\mu$m, 25$\mu$m,
60$\mu$m and 100$\mu$m.}
\end{figure}

Following Mattox et al. 1997 and Mattox et al. 2001, 
we find that the a-posteriori probability that 
the blazar we see in the EGRET error box is the correct identification is about 55\%.
The factors that we used for this calculation are the radio flux (2.1 Jy) and the spectral index
(+0.3), the 95\% error radius of the EGRET source (0$^{\circ}$.77), the distance of the radio source 
from the center of the EGRET circle (0$^{\circ}$.91), and the mean distance between radio sources 
that are at least as strong and at least as flat as this one ($\sim$15$^{\circ}$.4).
Although this value is not high enough to consider the association highly probable
(Mattox et al. 1997 require at least 70\%), it is nevertheless sufficiently secure
in the light of all the  other observational evidence.\\
At optical frequencies the source
is very faint and it is not reported in the USNO A2.0 
catalogue (Urban et al. 1997). However, spectral data over the range 5000-10000 $\AA$
are available 
from the work of Sowards-Emmerd et al. (2002) and can be used to
constrain the source properties.\\ We have also searched infrared
catalogues for a counterpart at these frequencies. There are several
2mass objects within the ROSAT error box, none of these however is
related to the radio object (see figure 4). Taking the lowest fluxes detected 
in the region as our sensitivity threshold, we
are able to infer an upper limit to the J,H,K bands respectively.
There is also an IRAS as well as an MSX (Midcourse Space
Experiment) source within the X-ray error box but its positional
uncertainty excludes coincidence with the radio object (see HEASARC Database).  Again we can
use the lowest far-infrared fluxes detected within 1 degree of our
source as our sensitivity limits to infer upper limits to the emission
in the 12-100 $\mu$m region. All these infrared upper limits are
reported in table 3. Although not simultaneously taken,
all available data are plotted in figure 5 in order to produce the Spectral
Energy Distribution (or SED) of 3EG J2027+3429; only the 2mass and MSX upper limits are excluded since they are
too close to the optical data.  In the widely adopted
scenario of blazars, a single population of high-energy electrons in a
relativistic jet radiate from the radio/FIR to the UV- soft X-ray by
the synchrotron process and at higher frequencies by inverse Compton
scattering of soft-target photons present either in the jet (synchrotron
self-Compton model), in the surrounding medium (external
Compton  model), or in both (Ghisellini et al. 1998 and references
therein). Therefore in the blazar SED, two peaks corresponding to the
synchrotron and inverse Compton components should be evident. This is
clearly compatible with the SED displayed in figure 5, where the first
peak is likely located in the millimeter far-infrared region while the
second most likely occurs in the soft gamma-ray range; furthermore
the X/gamma radiation completely dominates the radiative output.
Therefore, our object is consistent with being a low-frequency peaked
or red blazar. These spectral characteristics (synchrotron and Inverse
Compton peak at lower energies and dominance of the gamma-ray output)
are typical of flat spectrum radio quasars and so are fully compatible
with the classification of our proposed candidate.

\begin{table}
\begin{center}
\caption{Infrared upper limit measurements}
\begin{tabular}{lll}
\hline
    &  Wavelenght            & Upper limits             \\
    &   ($\mu$m)               & (erg cm$^{-2}$ s$^{-1}$)    \\
\hline
 1  & IRAS 12              & 5.06 $\times$ 10$^{-11}$ \\
 2  & IRAS 25             & 1.57 $\times$ 10$^{-11}$ \\
 3  & IRAS 60              & 2.92 $\times$ 10$^{-11}$  \\
 4  & IRAS 100             & 1.50 $\times$ 10$^{-10}$ \\
 5  & MSX  8               & 2.99 $\times$ 10$^{-11}$ \\
 6  & 2MASS (J) 1.25    & 7.19 $\times$ 10$^{-13}$ \\
 7  & 2MASS (K) 2.2     & 5.11 $\times$ 10$^{-13}$ \\
 8  & 2MASS (H) 1.65    & 9.46 $\times$ 10$^{-13}$ \\
\hline
\hline
\end{tabular}
\end{center}
\end{table}

\section{Conclusions}

From analysis of archived radio, infrared, optical and gamma-ray
data and from our own X-ray spectroscopy, we conclude that 3EG
J2027+3429 is a member of the blazar class of AGN. The X-ray source
spectrum from a few keV up to 100 keV is flat ($\gamma$=0.6-1.5) and highly
variable both in intensity and shape. The source SED confirms the
blazar nature of the source and furthermore it is compatible with its
classification as a flat spectrum radio quasar.  This identification
is interesting because the source is in the Galactic Plane and
it is the second one found in the Cygnus region. Recently,
Halpern et
al. (2001) identified a blazar-like radio source G74.87+1.22 as the
counterpart of the EGRET source 3EG J2016+3657 which is also on the Galactic
Plane. 
The probability of finding 2-3
EGRET blazars only 3 degrees from the Galactic
Equator can be estimated from the total number of relatively
well-identified blazars, 66, in the Third EGRET Catalog (Hartman et
al. 1999). This implies an expectation of around two blazars within the
zone -2.4$^{\circ}<$b$<$+2.4$^{\circ}$. Thus, we should not be surprised 
to have found another one.

\begin{acknowledgements}
We are grateful to Professor G.G.C. Palumbo for useful discussions and suggestions.
This research has made use of SAXDAS linearized and cleaned event files produced at
the BeppoSAX Science Data Center. It has also made use of data obtained from the High Energy Astrophysics Science Archive Research Center (HEASARC), provided by NASA's Goddard Space Flight Center.\\

\end{acknowledgements}


\begin{thebibliography}{}
\bibitem{} Campana, S., Parmar, A. N., Stella, L., 2001, A\&A, 372, 241
\bibitem{} Caraveo, P. 2002, the XXII Moriond Astrophysics Meeting "The Gamma-Ray Universe" 
           eds. A. Goldwurm, D. Neumann, and J. Tran Thanh Van, The GioiPublishers 
\bibitem{}Fiore, F., Guainazzi, M. \& Grandi, P. 1999,
Handbook for BeppoSAX NFI spectral analysis,
ftp://ftp.asdc.asi.it/pub/sax/doc/software\_docs/saxabc\_v1.2.ps.gz or
http://heasarc.gsfc.nasa.gov/docs/sax/abc/saxabc/
\bibitem{} Fossati, G., Maraschi, L., Celotti, A., Comastri, A., Ghisellini, G. 1998, MNRAS, 301, 451
\bibitem{} Frontera F., Costa E., Dal Fiume D., et al. 1997 A\&AS, 122, 357
\bibitem{} Ghisellini, G., Celotti, A., Fossati, G., Maraschi, L., Comastri, A. 1998, MNRAS, 301, 451
\bibitem{} Halpern, J. P., Eracleous, M., Mukherjee, R., Gotthelf, E. V.,  2001, ApJ, 551, 101
\bibitem{} Hartman, R. C., Bertsch, D. L., Bloom, S. D., Chen, A. W., Deines-Jones, P., et al. 1999, ApJS, 123, 79
\bibitem{} Mattox, J. R., Schachter, J., Molnar, L., Hartman, R. C., Patnaik, A. R. 1997, ApJ, 481, 95
\bibitem{} Mattox, J. R., Hartman, R. C., Reimer, O. 2001, ApJS, 135, 155
\bibitem{} McLaughlin, M. A., Mattox, J. R., Cordes, J. M., Thompson, D. J., APJ, 473, 763
\bibitem{} Merck, M.,Bertsch, D. L., Dingus, B. L., Esposito, J. A., Fichtel, C, et al. 1996, A\&A, 120, 465
\bibitem{} Mukherjee, R., Gotthelf, E. V., Halpern, J., Tavani, M. 2000, ApJ, 542, 740
\bibitem{} Mukherjee, R. 2001, Proceedings, volume 558. Edited by Felix A. Aharonian and Heinz J. Völk. 
           Published by American Institute of Physics, Melville, New York, ISBN 1-56396-990-4, 324 
\bibitem{} Romero, G. E., Benaglia, P., Torres, D. F., 1999, A\&A, 348, 868
\bibitem{} Sguera V., et al. 2003 in preparation
\bibitem{} Sowards-Emmerd, D., Romani, R. W., Michelson, P. F. 2002, American Astronomical Society Meeting 201, 11.18
\bibitem{} Urban, S., Corbin, T., Wycoff, G. 1997, AAS, 191, 5707 
\bibitem{} White, N. E, Giommi, P., Angelini, L. 1994, American Astronomical Society, 185th AAS Meeting, 
          N.41.11; Bulletin of the American Astronomical Society, Vol. 26, 1372
\bibitem{} Zhang, L., Zhang, Y. J., Cheng, K. S., 2000, A\&A, 357, 957




 

\end{thebibliography}
\end{document}